\documentclass[num-refs]{wiley-article}

\usepackage{siunitx}

\papertype{Research Article}
\paperfield{}

\title{Thickness-Driven Control of Room Temperature Ferrimagnetic Skyrmions and their Topological Hall signature in GdFe Single Layers}

\author[1,2\authfn{1}]{Saroj Kumar Mishra}
\author[2\authfn{2}]{Y.K. Takahashi}
\author[3\authfn{3}]{C. Malavika}
\author[3\authfn{3}]{Karthik V. Raman}
\author[1\authfn{1}]{Jyoti Ranjan Mohanty}

\affil[1]{Nanomagnetism and Microscopy Laboratory, Department of Physics, Indian Institute of Technology Hyderabad, Sangareddy 502285 Telangana, India}
\affil[2]{National Institute for Materials Science, Tsukuba, Ibaraki 305-0047, Japan}
\affil[3]{Tata Institute of Fundamental Research, Hyderabad, Telangana 500046, India}

\corraddress{Nanomagnetism and Microscopy Laboratory, Department of Physics, Indian Institute of Technology Hyderabad, Sangareddy 502285 Telangana, India}
\corremail{jmohanty@phy.iith.ac.in}

\fundinginfo{DST Nanomission (project no.: DST/NM/TUE/QM-4/2019-IIT-H); TIFR-Hyderabad from the Department of Atomic Energy, Government of India, Project Identification No. RTI 4007, and external funding from the ONRG Grant No. N62909-23-1-2049 }

\runningauthor{S. K. Mishra et al.}

\begin{document}

\begin{frontmatter}
\maketitle

\begin{abstract}
Magnetic skyrmions are nanoscale, topologically protected spin textures with exceptional potential for high-density data storage and energy-efficient computing. Controlling skyrmion characteristics, particularly size and density, is essential for advancing skyrmion-based device applications. Among various skyrmion-hosting systems, rare earth-transition metal ferrimagnets offer a promising platform due to their tunable magnetic properties and intrinsically low net magnetization. Despite this, the fundamental control of ferrimagnetic skyrmions in single-layer films remains unexplored. Here, we demonstrate a viable route for engineering room-temperature skyrmions in GdFe single layers through precise control of film thickness (60-80 nm). Thickness variation enables the systematic tuning of key magnetic parameters, including perpendicular magnetic anisotropy and saturation magnetization, thereby allowing precise control over skyrmion size and density. Magnetic force microscopy (MFM) reveals a clear thickness-dependent evolution of isolated skyrmion characteristics, where skyrmion size decreases while skyrmion density increases with increasing GdFe film thickness, in agreement with micromagnetic simulations. At the same time, magnetotransport measurements show a systematic enhancement in the topological Hall resistivity with thickness, further corroborating the increased skyrmion density observed in MFM. Scanning transmission electron microscopy reveals a compositional gradient across the film thickness, indicative of structural asymmetry and potential inversion-symmetry breaking, contributing to the emergence of a bulk Dzyaloshinskii–Moriya interaction. Notably, sub-60 nm skyrmions with high areal density are stabilized at room temperature. This work provides a viable route to tailor the properties of ferrimagnetic skyrmions in single-layer GdFe films, paving the way for the development of high-density ferrimagnetic skyrmionic devices.

% Please include a maximum of seven keywords
\end{abstract}
\end{frontmatter}

\section{INTRODUCTION}
Topologically protected, non-trivial spin textures have attracted significant interest due to their potential applications in next-generation logic and memory technologies \cite{fert2013skyrmions,everschor2018perspective}. Among these, nanoscale magnetic skyrmions represent the most extensively investigated spin configuration \cite{nagaosa2013topological}. In recent decades, they have emerged as a central topic within the spintronics community, owing to their distinctive topological properties, robust thermal stability, nanoscale dimensions, and the ultra-low critical current densities required for their manipulation \cite{fert2017magnetic}. These characteristics make skyrmions promising candidates for the development of high-density data storage systems \cite{hagemeister2015stability}, racetrack memory devices \cite{parkin2008magnetic}, spin-based logic circuits \cite{zhang2015magnetic}, transistor-analogous components \cite{zhang2015magnetics}, and neuromorphic computing platforms \cite{song2020skyrmion}. It is important to note that the key characteristics of skyrmions, such as their size, stability, and density are governed by intrinsic material parameters, including the saturation magnetization ($M_{s}$), exchange stiffness constant (A), Dzyaloshinskii–Moriya interaction (DMI) strength, magnetic anisotropy constant ($K_{u}$), and the film thickness. \cite{finocchio2016magnetic,bode2007chiral,heide2008dzyaloshinskii}. In magnetic thin films, DMI originates from the combined effects of strong spin–orbit coupling (SOC) and broken inversion symmetry, which may arise either from a non-centrosymmetric crystal lattice or from interfacial asymmetry that induces spin canting between neighboring atoms \cite{dzyaloshinsky1958thermodynamic,moriya1960anisotropic}. Apart from these, skyrmions are frequently investigated through Hall transport measurements, as their chiral spin textures give rise to an additional contribution to the Hall resistivity, known as the topological Hall effect (THE) \cite{lee2009unusual,kanazawa2011large}. THE serves as a sensitive and practical probe for identifying the presence of skyrmions within magnetic systems. Initially observed as a Hall response arising from spin chirality in manganites and frustrated ferromagnets, later THE was subsequently attributed to skyrmionic spin textures in B20-type bulk materials \cite{muhlbauer2009skyrmion}.

Stabilization of magnetic skyrmions has been demonstrated in multilayer stacks such as Ir/Fe/Co/Pt \cite{soumyanarayanan2017tunable}, Pt/Co/Ta \cite{woo2016observation}, Co/Pd \cite{pollard2017observation}, and recently in CoPt single layers \cite{erickson2024room}. However, in ferromagnetic materials, several technological challenges hinder the development of skyrmion-based memory and logic devices. For instance, the skyrmion Hall effect (SkHE) presents challenges for achieving controlled, unidirectional motion along racetrack architectures. \cite{fert2013skyrmions,litzius2017skyrmion} Furthermore, substantial reduction of skyrmion sizes is necessary for high-density device integration; however, achieving thermal stability of such nanoscale skyrmions at room temperature (RT) remains a critical issue. Amorphous rare earth–transition metal (RE–TM) ferrimagnets (FiMs), such as Gd–FeCo, Co–Gd, Tb–Fe, and Gd-Fe, have emerged as promising candidates for addressing these challenges. Recent studies have demonstrated the formation of small-sized skyrmions exhibiting reduced SkHE in multilayer systems such as Pt/GdFeCo/MgO \cite{woo2018current}, Pt/GdCo/Ta \cite{caretta2018fast}, and Pt/TbFe/Ta \cite{xu2023systematic}. In these structures, the interfacial DMI, which plays a crucial role in stabilizing skyrmions, typically decreases with increasing magnetic layer thickness, thereby reducing interfacial magnetic anisotropy. \cite{yang2015anatomy,belmeguenai2018thickness} Increasing the magnetic layer thickness is often necessary to improve thermal stability \cite{buttner2018theory,siemens2016minimal}, which leads to a reduction in interfacial DMI. 

To overcome the limitations of interface-dependent DMI in multilayer systems, a promising alternative is to use single-layer materials where bulk DMI governs skyrmion stability. However, materials exhibiting bulk DMI are relatively rare, and their DMI strength is inherently fixed by crystal symmetry, making it difficult to tune. In amorphous alloys, the absence of long-range order generally prevents the emergence of DMI due to the lack of a preferred orientation. Contrary to this expectation, prior studies have reported measurable bulk DMI in amorphous Gd-FeCo alloys, attributed to atomic-scale compositional inhomogeneities between RE and TM elements and the direct evidence for the presence of SOC \cite{kim2019bulk,krishnia2021spin}. In contrast, we recently reported a systematic, thickness-dependent investigation of interface-independent DMI in a GdFe single-layer film using analytical domain models and micromagnetic simulations \cite{kumar2024engineering}. Amorphous RE–TM films offer several intrinsic features favorable for skyrmion formation, such as their intrinsic non-collinear spin configurations arising from competing exchange interactions between antiparallel RE and TM sublattices\cite{buttner2018theory}. Their intrinsic magnetic properties, such as $M_{s}$ and $K_{u}$, can be independently tuned by altering the composition of RE and TM sublattices \cite{xu2023systematic,mallick2024driving}. This flexibility allows precise control over exchange, DMI, and dipolar energies that collectively stabilize small skyrmions \cite{harris1992structural,hansen1989magnetic,heigl2021dipolar}. Moreover, the domain walls often exhibit strong demagnetizing fields, providing a nucleation platform for skyrmions under modest external magnetic fields \cite{legrand2018hybrid}. In addition, their amorphous nature minimizes crystalline defects and pinning, enabling the development of high-speed and stable skyrmion-based devices \cite{mallick2024driving,kim2017fast}. These collective advantages offer a route to stabilize skyrmions at RT without the complexity of multilayer heterostructures, interface-related issues, and the need for fabrication, while enhancing compatibility with spintronic devices \cite{kim2022ferrimagnetic}. Nevertheless, the direct visualization and characterization of such chiral spin textures and their topological signatures in amorphous single-layer RE–TM films remain unexplored. This gap in understanding serves as the primary motivation for our present study.

This work demonstrates the formation of RT FiM skyrmions across varying thicknesses in single-layer Gd-Fe films. Using scanning transmission electron microscopy, we identify an asymmetric distribution of elemental composition, which likely contributes to the broken inversion symmetry within the FiM layer. The RT skyrmions were investigated comprehensively through SQUID magnetometry, magnetic force microscopy (MFM), and magneto-transport measurements. Isolated skyrmions were directly imaged via MFM and found to correlate with a field-dependent THE, providing clear evidence of nontrivial topology. The evolution of domain structures and the nucleation of isolated skyrmions were systematically studied over a wide range of magnetic fields at RT. By tuning the FiM layer thickness and the external magnetic field, we achieved control over skyrmion stability, including modulation of skyrmion density ($\eta_s$) and diameter ($d_s$). Micromagnetic simulations were conducted to gain deeper insight into the mechanisms underlying skyrmion formation, yielding results consistent with experimental observations. These findings expand the range of viable material systems capable of hosting compact and high-density RT magnetic skyrmions in single-layer systems, independent of interfacial engineering.

\section{RESULTS AND DISCUSSION}
\begin{figure*}[h]
\centering
\includegraphics[width = 6in ,height=4.2in]{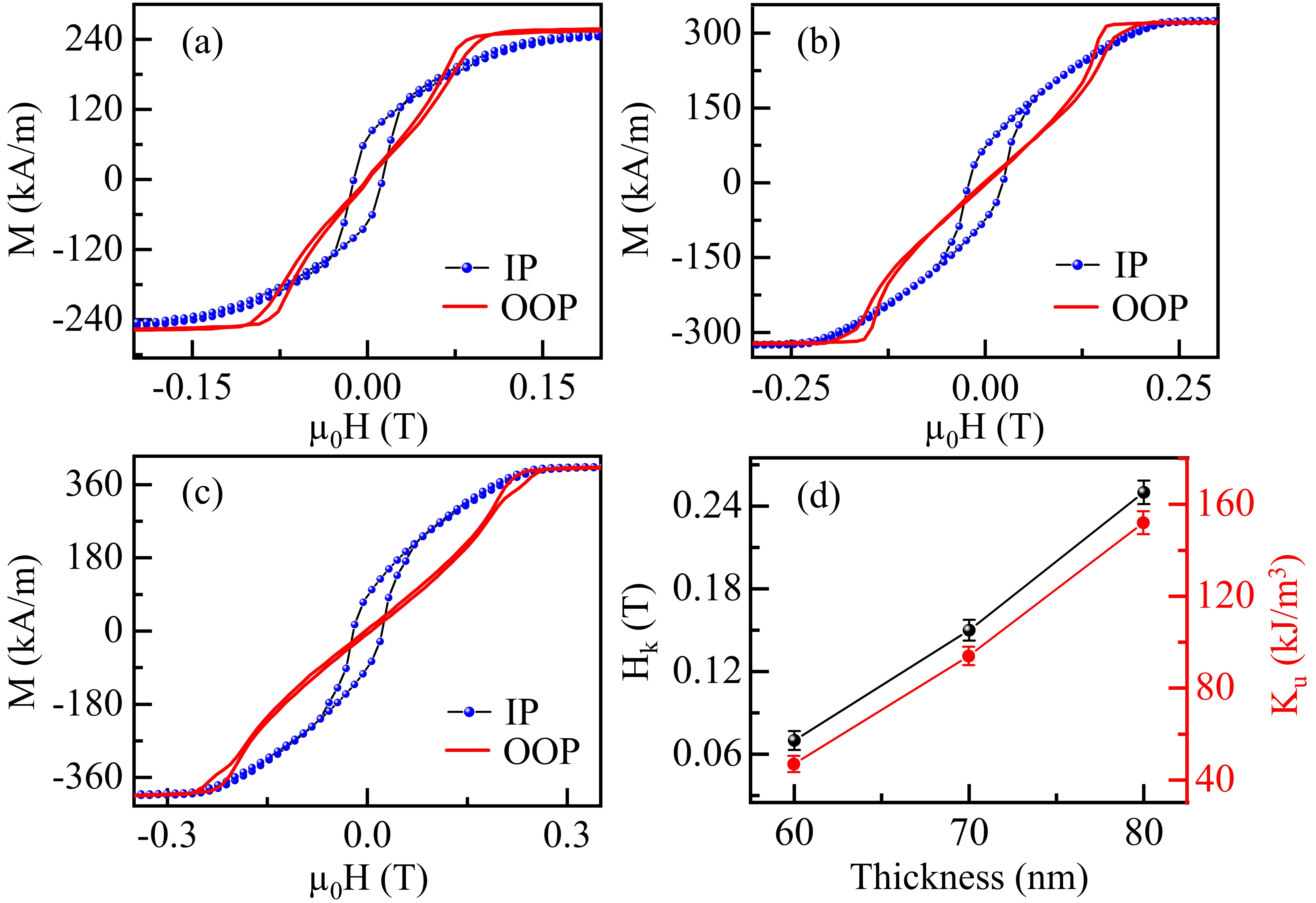}
\caption{In-plane and out-of-plane magnetization hysteresis loops for GdFe films with thicknesses of (a) 60 nm, (b) 70 nm, and (c) 80 nm. (d) Extracted anisotropy field ($H_{k}$) and uniaxial magnetic anisotropy ($k_{u}$) as a function of film thickness.}
\end{figure*}

GdFe single-layer films with thicknesses of 60, 70, and 80 nm were fabricated via electron-beam evaporation (see Experimental Section). The out-of-plane (OOP) and in-plane (IP) magnetization–field hysteresis loops of the films were measured at room temperature (RT) using superconducting quantum interference device (SQUID) magnetometry. Figure 1 (a,b,c) presents the OOP and IP loops for 60, 70, and 80 nm thicknesses, respectively. All key magnetic parameters exhibit a clear thickness dependence: the saturation field ($\mu_{0}H_{s}$) increases from 0.15 T to 0.3 T, while the saturation magnetization ($M_{s}$) increases from 240 kA/m to 370 kA/m with increasing thickness. The OOP loops show a low-remanence, slanted shape characteristic, indicating the presence of labyrinth domains and skyrmionic signatures in the samples \cite{woo2016observation,soumyanarayanan2017tunable,zhang2018determination,bhatti2023enhancement}. The nearly linear growth of magnetization with field and the narrow opening near the saturation field are consistent with the reversible expansion of stripe domains parallel to the applied field. At higher fields, the loop opening reflects irreversible processes such as the nucleation and annihilation of cylindrical spin textures \cite{malozemoff1979magnetic}. Analysis of the OOP and IP loops reveals that the films exhibit weak PMA with a very low anisotropy field ($\mu_{0}H_{k}$) increasing only from 0.06 T to 0.24 T with thickness. Such low $\mu_{0}H_{k}$  values indicate that the magnetic configuration is highly susceptible to IP fields. Previous studies have demonstrated that skyrmion phases can be stabilized by tuning the PMA, with their emergence typically favored in weak-PMA regimes \cite{yu2016room,zhang2018creation}. The corresponding magnetic anisotropy energy density ($K_{u} = \frac{ \mu_{0}H_{k}M_{s}}{2} + \frac{1}{2} \mu_{0}M_{s}^2$) is calculated which increases from 50 to 160 $kJ/m^{3}$ across the thickness range (see Figure 1d), indicating tunable PMA in our films. Such tunability is consistent with earlier studies of RE–TM ferrimagnets \cite{mishra2017anomalous,talapatra2019observation,mishra2025exploring}. 

\subsection{Correlating Topological Hall Response to Real-Space Imaging of Ferrimagnetic Skyrmions.}
Magneto-transport measurements were carried out at RT to probe the presence of chiral spin textures in GdFe films of varying thickness. Hall transport is a powerful probe for chiral spin textures, such as skyrmions, as their nontrivial topology contributes an additional Hall signal, the topological Hall effect (THE), which coexists with the ordinary Hall effect (OHE) and the anomalous Hall effect (AHE). When a charge current flows through non-collinear spin textures, the resulting real-space Berry curvature produces an emergent magnetic field that deflects conduction electrons transverse to their trajectory,  giving rise to the THE.\cite{lee2009unusual,kanazawa2011large} The topological hall resistivity is expressed as $\Delta \rho_{xy}(H) = PR_0B_{eff} = PR_0n_{sk}\Phi_0$, where $0 < P < 1$ denotes the transport spin-polarization of the charge carriers, $R_0$ is the effective charge density to the THE, and $B_{eff}$ is the effective magnetic flux density arising from the presence of skyrmions \cite{raju2021colossal,neubauer2009topological}. This effective magnetic flux density is defined as $B_{eff} = n_{sk}\Phi_0$, where $n_{sk}$ is the skyrmion density and $\Phi_0$ is the magnetic flux quantum. Experimentally, $\Delta \rho_{xy}(H)$ was extracted using the relation $\Delta \rho_{xy}(H)$ = $\rho_{xy}(H)$ - $\rho_{xy}^{Fit}(H)$, where $\rho_{xy}(H)$ is the total Hall resistivity. Here, $\rho_{xy}^{Fit}(H)$ = $R_{0}(H) + R_{s}M(H)$, where $R_{0}$ is the ordinary Hall coefficient and $R_{s}$ accounts for the anomalous Hall effect arising from the intrinsic momentum-space Berry-curvature mechanism and extrinsic mechanisms such as skew scattering, side-jump processes. After carefully separating these conventional Hall contributions, the residual signal $\Delta \rho_{xy}(H)$ (green circles in Figures 2a, 3a, and 4a) was obtained by subtracting the fitted background $\rho_{xy}^{Fit}(H)$ (red curves) from the experimentally measured Hall resistivity $\rho_{xy}(H)$ (black curves). The $\rho_{xy}(H)$ measurements were carried out using a PPMS, while the corresponding magnetization $M(H)$ was determined by SQUID magnetometry (see Experimental Section). A clear deviation between the measured and fitted curves emerges near the opening of the hysteresis loop for all the film thicknesses, and this difference manifests as a characteristic peak and dip-like feature in $\Delta \rho_{xy}(H)$. This behavior is a hallmark of the THE and may arise from non-collinear spin textures, such as skyrmions stabilized by strong bulk DMI, which in turn originates from the anisotropic antiferromagnetic coupling between the Gd (4f) and Fe (3d) sublattices in the GdFe thin film. 
\begin{figure*}[h]
\centering
\includegraphics[width = 5.5in ,height=3.8in]{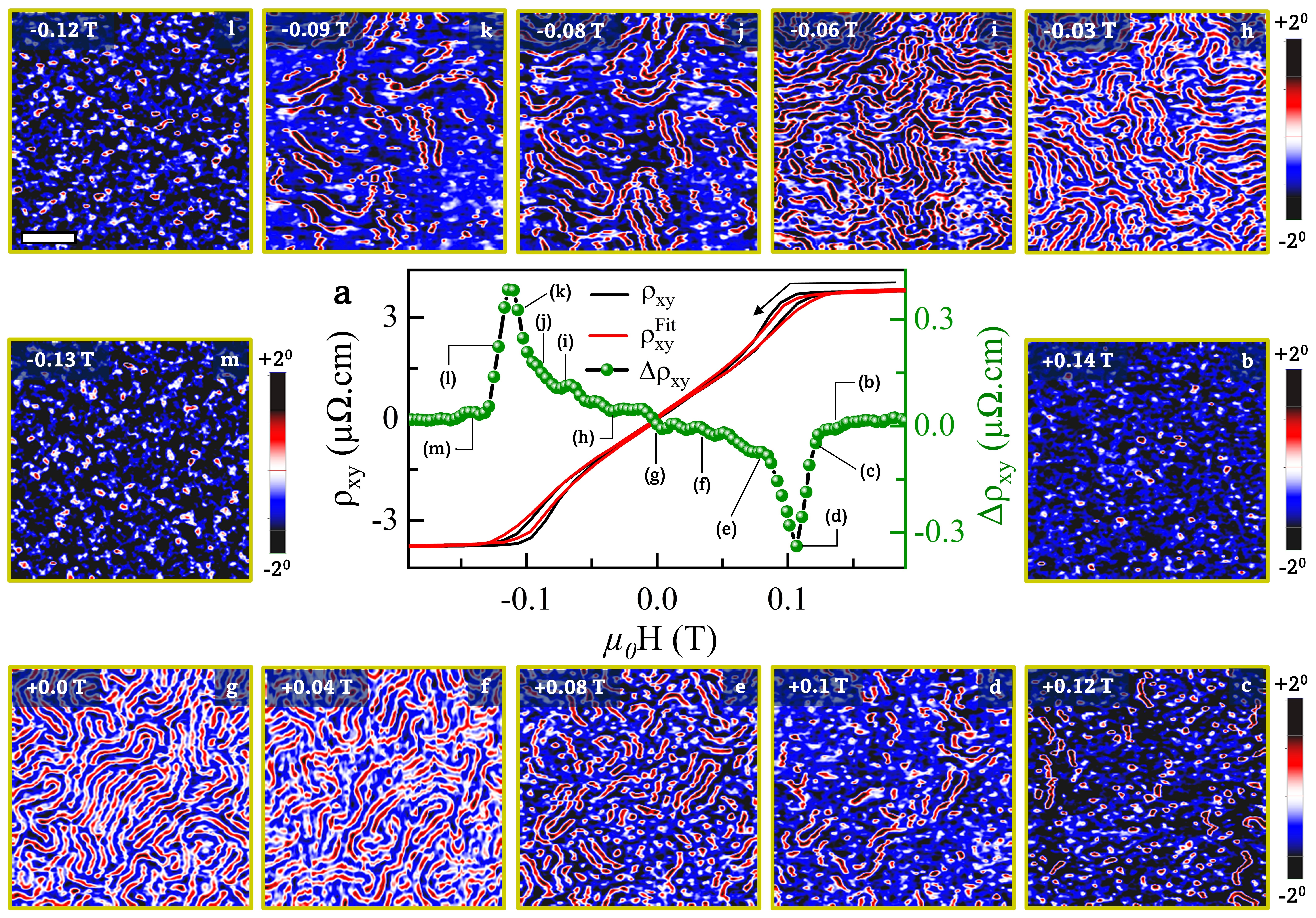}
\caption{Room temperature MFM imaging of magnetic spin textures and THE as a function of applied field (H) in a 60 nm GdFe single layer. $\boldsymbol{a}$ represents topological Hall resistivity ($\Delta \rho_{xy}(H)$) (green circles) obtained by subtracting the fitted ordinary and anomalous Hall contributions ($\rho_{xy}^{Fit}(H)$, red curve) from the total Hall resistivity ($\rho_{xy}(H)$, black curve) ($\Delta \rho_{xy}(H)$ = $\rho_{xy}(H)$ - $\rho_{xy}^{Fit}(H)$). The black arrow indicates the field sweep direction for $\Delta \rho_{xy}$ and MFM, while $\rho_{xy}$ and $\rho_{xy}^{Fit}$ are shown for both sweep directions. $\boldsymbol{b-m}$ Corresponding MFM images at selected OOP magnetic fields, as marked on the topological Hall curve and top-left corners of each panel. All MFM images share the same scale bar shown in panel l, corresponding to 1 $\mu m$.}
\end{figure*}
To assess whether the applied measurement current could influence the stability of the skyrmion textures, we estimated the corresponding current densities for the Hall measurements performed with a dc current of $200 \mu A$. For the 60 nm, 70 nm, and 80 nm thick GdFe films with a sample width of 4 mm, the resulting current densities are approximately $8.3 \times 10^5 $, $7.1 \times 10^5 $, and $6.3 \times 10^5 $ $A/m^2$, respectively. These values are several orders of magnitude lower than the typical current densities required to induce current-driven skyrmion motion or deformation ($\sim 10^{10}-10^{12}$) \cite{woo2018current,caretta2018fast}. Therefore, the applied measurement current does not introduce any current-induced effects, and the observed topological Hall signals faithfully reflect the intrinsic skyrmion properties of the GdFe films.

\begin{figure*}[ht]
\centering
\includegraphics[width = 5.5in ,height=3.8in]{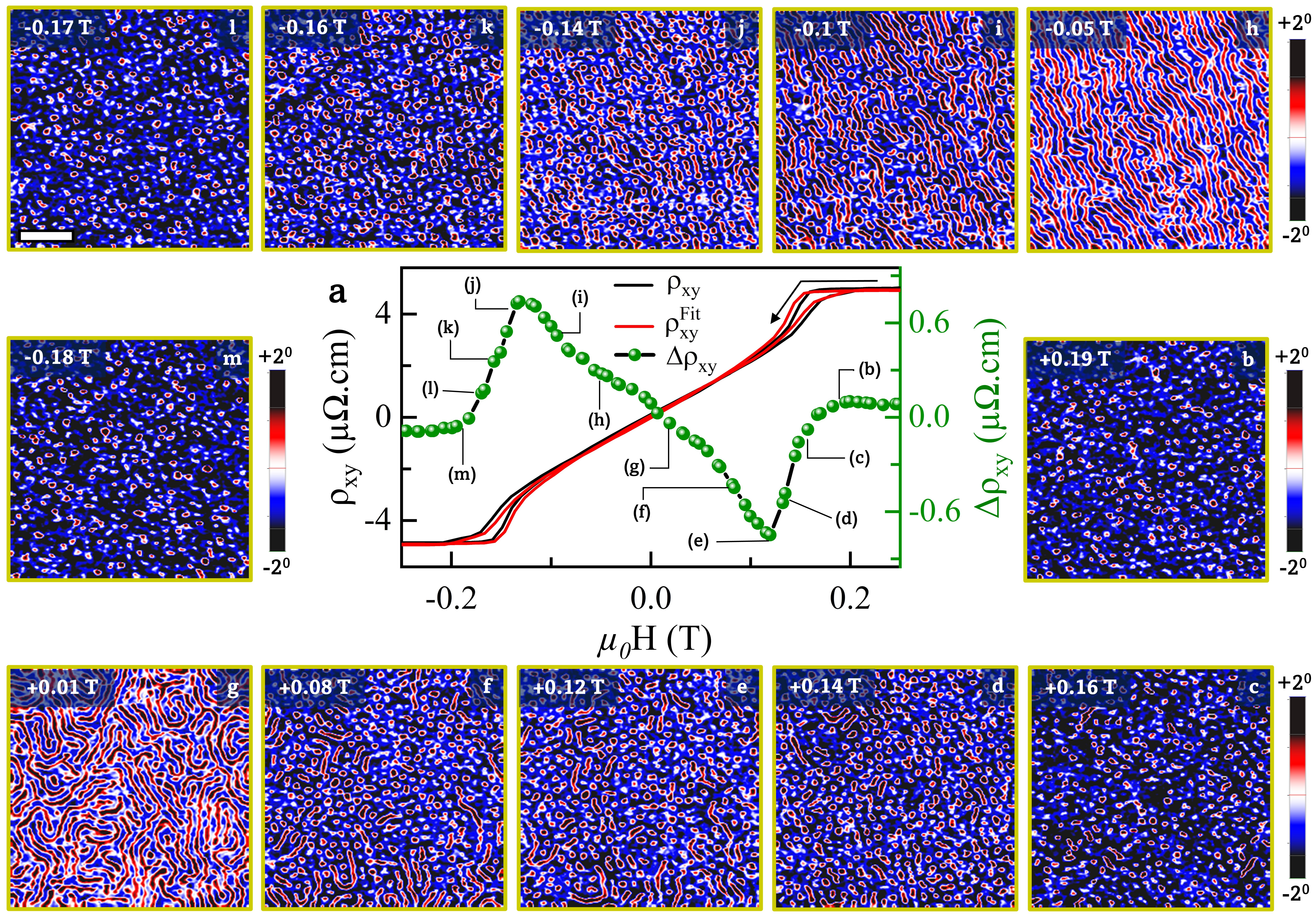}
\caption{Room temperature MFM imaging of magnetic spin textures and THE as a function of applied field (H) in a 70 nm GdFe single layer. $\boldsymbol{a}$ represents topological Hall resistivity ($\Delta \rho_{xy}(H)$) (green circles) obtained by subtracting the fitted ordinary and anomalous Hall contributions ($\rho_{xy}^{Fit}(H)$, red curve) from the total Hall resistivity ($\rho_{xy}(H)$, black curve) ($\Delta \rho_{xy}(H)$ = $\rho_{xy}(H)$ - $\rho_{xy}^{Fit}(H)$). The black arrow indicates the field sweep direction for $\Delta \rho_{xy}$ and MFM, while $\rho_{xy}$ and $\rho_{xy}^{Fit}$ are shown for both sweep directions. $\boldsymbol{b-m}$ Corresponding MFM images at selected OOP magnetic fields, as marked on the topological Hall curve and top-left corners of each panel. All MFM images share the same scale bar shown in panel l, corresponding to 1 $\mu m$.}
\end{figure*}

To elucidate the microscopic origin of the characteristic double-peak features with opposite sign (i.e., both a peak and a dip) in the topological Hall response, we performed magnetic force microscopy (MFM) (see Experimental Section) imaging on the GdFe films at RT. The black arrows in Figures 2a, 3a, and 4a denote the field-sweep direction for both $\Delta\rho_{xy}$ and MFM, while Figures 2b–m, 3b–m, and 4b–m display representative MFM images at selected OOP magnetic fields, as marked on the corresponding Hall curves. 
\begin{figure*}[ht]
\centering
\includegraphics[width = 5.5in ,height=3.8in]{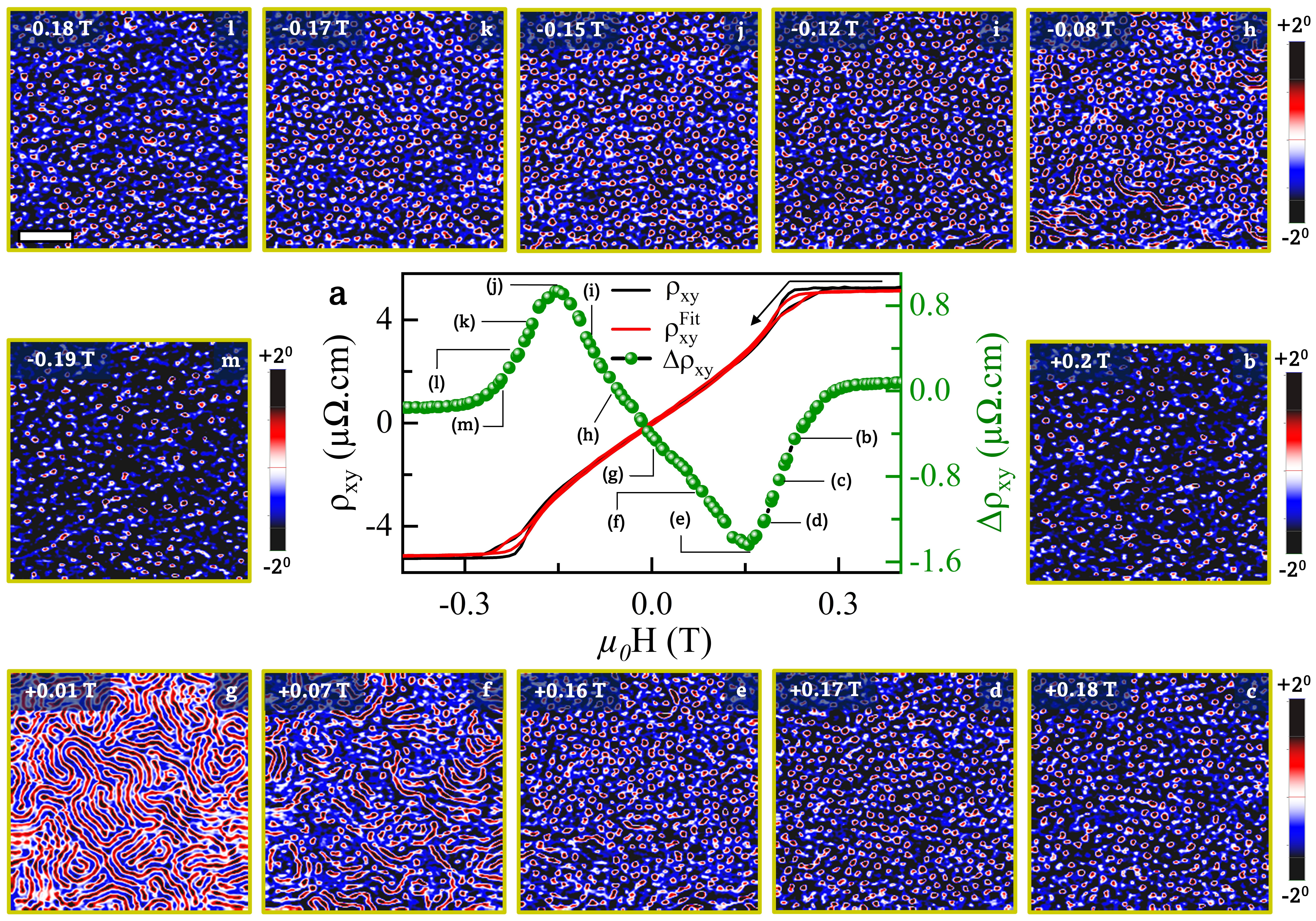}
\caption{Room temperature MFM imaging of magnetic spin textures and THE as a function of applied field (H) in a 80 nm GdFe single layer. $\boldsymbol{a}$ represents topological Hall resistivity ($\Delta \rho_{xy}(H)$) (green circles) obtained by subtracting the fitted ordinary and anomalous Hall contributions ($\rho_{xy}^{Fit}(H)$, red curve) from the total Hall resistivity ($\rho_{xy}(H)$, black curve) ($\Delta \rho_{xy}(H)$ = $\rho_{xy}(H)$ - $\rho_{xy}^{Fit}(H)$). The black arrow indicates the field sweep direction for $\Delta \rho_{xy}$ and MFM, while $\rho_{xy}$ and $\rho_{xy}^{Fit}$ are shown for both sweep directions. $\boldsymbol{b-m}$ Corresponding MFM images at selected OOP magnetic fields, as marked on the topological Hall curve and top-left corners of each panel. All MFM images share the same scale bar shown in panel l, corresponding to 1 $\mu m$.}
\end{figure*}
At low fields, the $\Delta\rho_{xy}$ signal weakens, reflecting the predominance of labyrinthine domains (Figures 2g, 3g, and 4g), consistent with the small remanence observed in the hysteresis loops (Figure 1). These labyrinthine textures arise from the competition among exchange, anisotropy, and dipolar interactions. Upon increasing the field, the labyrinth domains gradually evolve into a mixed phase of elongated stripes and circular spin textures, driven by the Zeeman energy, which aligns the magnetization with the external field. The elongated stripe domains diminished as the field approached saturation, leaving only randomly distributed circular spin textures, as shown in Figures 2b, m, 3b, m, and 4b, m. Intermediate-field MFM scans verified that this transformation occurred continuously across all investigated film thicknesses. The complete field-dependent evolution of skyrmion from positive to negative saturation is provided for the 60 nm, 70 nm, and 80 nm GdFe films in Supplementary Movies 1–3. Importantly, these circular textures persisted over a wide field range once nucleated, reflecting their intrinsic stability. At the same time, all samples exhibited pronounced double-peak features in the $\Delta\rho_{xy}$. The concurrent observation of stable circular textures in MFM and characteristic THE features in transport rules out trivial bubble domains, unambiguously identifying the observed textures as skyrmions with a well-defined chirality.

To understand the origin of the observed double-peak features in the $\Delta \rho_{xy}$ curve, we recall that, $\Delta \rho_{xy}$, is proportional to the average topological charge (skyrmion number) \cite{sivakumar2020topological}, i.e., $\Delta \rho_{xy} = R_T \langle Q_{sk} \rangle$, where $Q_{sk}$ denotes the sum of the skyrmion numbers of all topological objects present, and $R_T$ is the topological Hall coefficient determined by the carrier density and spin polarization of the conduction electrons. The skyrmion topological charge is defined as $\langle Q_{sk} \rangle = V \cdot P$, where $V$ is the vorticity and $P$ is the core polarity \cite{nagaosa2013topological,jefremovas2025role}. For skyrmions, $V = +1$, whereas the polarity $P$ corresponds to the orientation of the skyrmion core magnetization relative to the film normal: $P = +1 (-1)$ for a $+M_s (-M_s)$ magnetized sample. Generally, a sign reversal of $R_T$ is expected only with temperature changes, since varying the external magnetic field does not alter the carrier densities. So, the sign change in $R_T$ can only vary with temperature due to alterations in the electronic band structure and carrier concentration \cite{sivakumar2020topological}. Since all measurements here were conducted at RT, the sign changes observed in $\Delta \rho_{xy}$ arise from polarity switching of the skyrmion cores. Thus, the double-peak feature with opposite signs reflects the coexistence of two distinct skyrmion phases, stabilized sequentially during the magnetic-field sweep as the skyrmion core polarity reverses. A similar observation has recently been reported in both Pt/Co/Ta and Pt/Co/Re systems \cite{ojha2025tailoring}. Importantly, the $\Delta \rho_{xy}$ signal associated with these non-collinear spin textures is confined to the field regime where skyrmions are present and vanishes once the magnetization saturates, which would not be expected for conventional anomalous Hall contributions. This distinction has previously been used to separate genuine topological Hall signals from artifacts arising from multi-component AHE contributions \cite{tai2022distinguishing}. Furthermore, side-jump and skew-scattering mechanisms are symmetric with respect to field reversal and therefore cannot account for the observed field-dependent sign reversals and double-peak feature in $\Delta \rho_{xy}$, supporting its topological origin.

Generally, in systems with weak PMA, non-collinear spin textures such as twisted domain walls or magnetic bubbles may form and, in principle, contribute to transverse transport signals. In the present GdFe films, MFM imaging reveals a mixed magnetic phase near the onset of the THE, consisting of elongated stripe domains, magnetic bubbles, and skyrmions. However, as the OOP magnetic field increases, these stripe-like and bubble-like textures are progressively annihilated, leaving a magnetic state dominated by isolated, well-separated skyrmions. Notably, the maximum THE signal coincides with this skyrmion-dominated regime. Since magnetic bubbles are topologically trivial ($Q_{sk}$=0) and are not expected to generate a robust topological Hall response, while skyrmions carry a finite topological charge, the observed THE is attributed primarily to skyrmions. Although minor contributions from other non-collinear textures at low fields cannot be completely excluded, their influence is negligible in the high-field regime where the THE reaches its maximum.

Furthermore, the extracted $\Delta \rho_{xy}$ shows a clear thickness dependence. The average magnitude of $\Delta \rho_{xy}$ signal systematically increases with film thickness, from $\sim$ 0.36 $\mu\Omega.cm$ in the 60 nm film, to $\sim$ 0.73 $\mu\Omega.cm$ in the 70 nm film, and up to $\sim$ 1.18 $\mu\Omega.cm$ in the 80 nm film. This trend correlates directly with the increase in skyrmion density, $n_{sk}$, revealed by MFM, consistent with the scaling relation $\Delta \rho_{xy} \propto n_{sk}$. Notably, there is an asymmetry in the magnitude of $\Delta \rho_{xy}$ signals during field sweep, i.e., $\Delta \rho_{xy}(+H) \neq \Delta \rho_{xy}(-H)$, which increases with increasing thickness. Such behavior can be correlated with the intrinsic FiM nature of GdFe. Specifically, the anisotropic exchange coupling between the non-collinear Gd (4f) and Fe (3d) sublattices generates inequivalent energy landscapes for skyrmion stabilization under opposite field polarities. In addition, the traversal of conduction electrons through the non-collinear spin textures can lead to asymmetric spin-dependent scattering \cite{denisov2020theory}, arising from the distinct coupling strengths to the two sublattices. 

\subsection{Depth Resolved Microstructural Analysis}
To explore the possible origin of inversion-symmetry breaking in amorphous GdFe single layers, we performed depth-resolved microstructural analysis using scanning transmission electron microscopy (STEM) combined with energy-dispersive X-ray spectroscopy (EDS) (see Experimental Section). High-angle annular dark-field (HAADF) STEM images of 60, 70, and 80 nm GdFe films (Figure 5a–c) show that the GdFe layer appears brighter than the Cr capping layer due to Z-contrast, and exhibits the characteristic uniform amorphous contrast across its entire thickness, excluding the possibility of a non-centrosymmetric crystalline structure. To investigate elemental distributions, STEM–EDS mapping was performed. Elemental maps of Gd, Fe, and Cr (Figure 5d–l), along with composite overlays (Figure 5m–o), reveal sharp interfaces with no detectable interdiffusion between the Si substrate, the GdFe layer, and the Cr cap. However, such 2D mapping cannot unambiguously resolve compositional variations across the film thickness (vertical) because it integrates information along the electron beam (horizontal) direction. To overcome this limitation, we extracted laterally averaged compositional line profiles along the film normal (z-axis) (Figure 5p–r) for 60, 70, and 80 nm, respectively. A pronounced composition gradient in Fe and Gd is observed for all thicknesses. Quantitative analysis of the Gd-to-Fe composition ratios (Figure 5s–u) reveals clear deviations from uniformity, with linear fits yielding slope values ($\beta$) that quantify the magnitude of the gradient. The extracted $\beta$ decreases with increasing film thickness (Figure 5v), indicating a systematic reduction in structural inversion asymmetry.

\begin{figure*}[htbp]
\centering
\includegraphics[width = 5.5in ,height=5.1in]{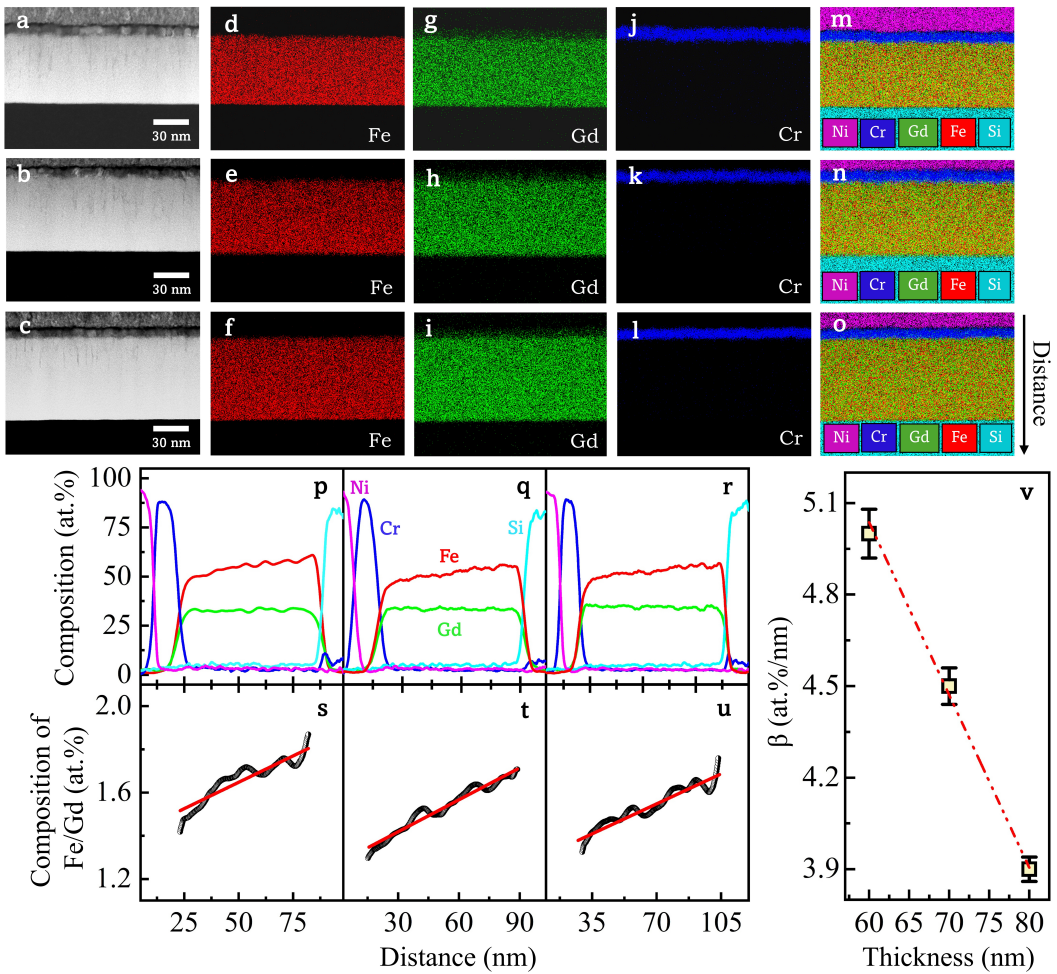}
\caption{Depth-resolved microstructural analysis of GdFe films via STEM and EDS. High-angle annular dark field STEM images of GdFe films with thicknesses of (a) 60 nm, (b) 70 nm, and (c) 80 nm. (d–f), Corresponding elemental maps of the Fe K-edge. (g-i), Elemental maps of the Gd L-edge. (j-l), Elemental maps of the Cr K-edge. (m-o), Composite elemental maps showing the spatial distribution of Fe, Gd, Cr, Si, and Ni for each thickness. (p-r), Laterally averaged line profiles along the z-axis direction for all elemental components in the 60, 70, and 80 nm GdFe films, respectively. (s-u) Corresponding averaged line profiles of the Fe/Gd composition as a function of z-position. Solid lines represent linear fits, with the slope denoted by $\beta$. (v) Extracted slope $\beta$ plotted as a function of film thickness.}
\end{figure*}

These findings demonstrate that this elemental inhomogeneity induces structural inversion asymmetry in the amorphous GdFe layers. Such asymmetry, in combination with SOC, can generate a net bulk DMI, thereby facilitating the stabilization of skyrmionic spin textures. We also note that the 3 nm Cr capping layer is too thin to sustain magnetic ordering or significant spin-orbit-driven interfacial DMI, as ultrathin Cr films remain nonmagnetic and lack the spin-density-wave ordering present in thicker Cr layers \cite{meersschaut1995spin,zabel1999magnetism}. This suggests that the observed DMI originates predominantly from the internal structural asymmetry of the GdFe film, rather than from the GdFe/Cr interface. A comparable thickness-dependent compositional gradient in GdFeCo single-layer films has been reported previously \cite{kim2019bulk,krishnia2021spin}. We attribute the observed compositional gradient to the combined effects of the distinct vapor pressures of Gd and Fe, along with slower deposition rates and lower sticking coefficients during e-beam growth. All the films were deposited using a $Gd_{50}Fe_{50}$ alloy source under identical growth conditions. Given that Fe possesses a higher vapor pressure and correspondingly higher evaporation rate than Gd under the same conditions. As a result, Fe atoms are evaporated more readily and rapidly into the growing film, whereas Gd atoms accumulate at a slower rate. This imbalance can lead to a vertical composition gradient along the growth direction. As thickness increases, the films become more uniform, resulting in a reduction in the compositional gradient.

\subsection{Experimental Validation of Skyrmion Dynamics via Micromagnetic Simulations} 
To further verify the experimentally observed $\Delta \rho_{xy}$ behavior and to gain deeper insights into the underlying skyrmion spin textures, we performed micromagnetic simulations using $Mumax^3$ for all GdFe film thicknesses \cite{vansteenkiste2014design}. The input magnetic parameters used in the simulations were obtained from the SQUID magnetometry measurements (see Experimental Section). Bulk DMI is the primary stabilizing mechanism for skyrmions in single-layer films, and its magnitude is known to increase with film thickness \cite{kim2019bulk}. First, we attempted simulations using a DMI strength of 0.35 $mJ/m^2$ for the 60 nm GdFe film, but they failed to reproduce the experimentally observed zero-field labyrinth domain state, as shown in Figure S1. When the bulk DMI value of 0.4 $mJ/m^2$ was used, simulated field-induced domains closely matched the MFM results, as shown in Figure 6a. When we fix the DMI value of 0.4 $mJ/m^2$ to thicker films of 70 nm and 80 nm, the simulated domain morphology and skyrmion characteristics no longer matched the experimental observations, as shown in Figure S2. For this, we systematically varied the bulk DMI strength from 0.4 to 0.6 $mJ/m^2$ as the thickness increased. These values are very close to the previously reported bulk DMI strengths \cite{kim2019bulk}. The simulation starts with the magnetization reversal process by sweeping the OOP magnetic field from positive to negative saturation with a step size of 0.002 T, while capturing domain images at each field step. In addition, to clarify the origin of the experimentally observed $\Delta \rho_{xy}$, we quantified the field-dependent evolution of the average skyrmion number, $\langle Q_{sk} \rangle$, at each applied field using the “ext-topologicalcharge” function in $Mumax^3$. Figure 6 summarizes the simulated OOP hysteresis loops, the corresponding $\langle Q_{sk} \rangle$ curves, and the domain configurations for the 60 nm, 70 nm, and 80 nm films.

\begin{figure*}[htbp]
\centering
\includegraphics[width = 5.5in ,height=5.3in]{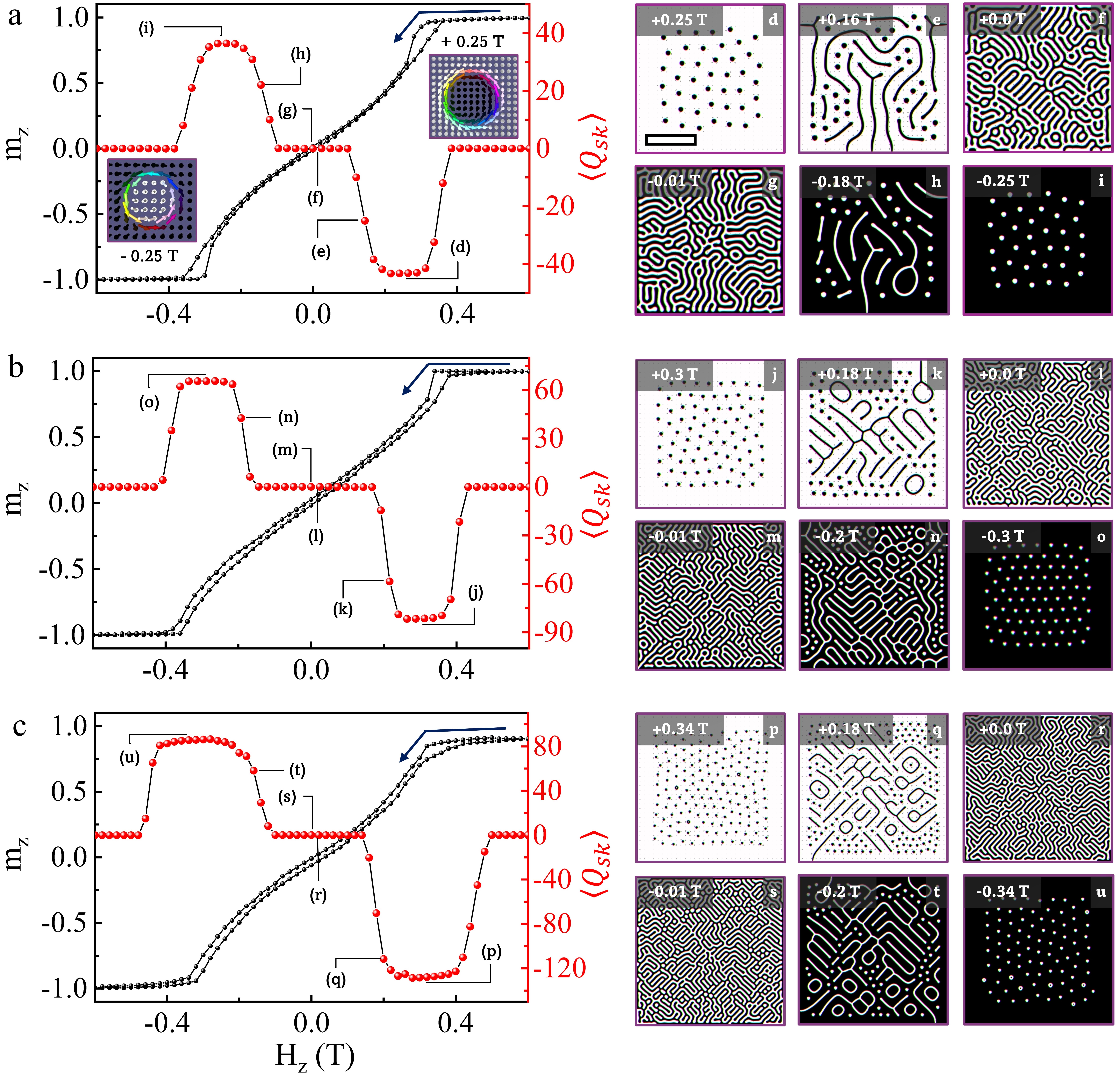}
\caption{Simulated out-of-plane hysteresis loops and corresponding average skyrmion number $\langle Q_{sk} \rangle$ for GdFe films with thicknesses of (a) 60 nm, (b) 70 nm, and (c) 80 nm. The blue arrow indicates the field sweep direction. (d-i), (j-o), and (p-u) display the simulated domain configurations for 60, 70, and 80 nm films, respectively, at selected OOP magnetic fields, as marked on the $\langle Q_{sk} \rangle$ curve and top-left corners of each panel. The inset in (a) shows the Bloch skyrmions with opposite polarity obtained at 0.25 and -0.25 T. All domain images share the same scale bar shown in panel d, corresponding to 0.5 $\mu m$.} 
\end{figure*}

A notable difference between the experiment and the simulation is the slightly higher saturation fields observed in the simulation. This discrepancy is expected, as experimental magnetization reversal can be influenced by imperfections, such as pinning centers and local inhomogeneities, which are absent in the idealized simulated films. Nevertheless, the overall loop characteristics, including the systematic increase of saturation field from 0.4 T to 0.5 T with increasing thickness and the reduction of loop opening near saturation, are in excellent agreement with experimental trends. Furthermore, the simulated domain evolution (Figure 6) closely reproduces the MFM observations (Figures 2-4). As the field decreases from positive saturation, isolated skyrmions with$\langle Q_{sk} \rangle$ = -1 emerge (Figure 6(d,j,p)). With further field reduction, these skyrmions expand into elongated stripe domains. At zero field, a labyrinth domain pattern develops (Figure 6(f,l,r)), corresponding to maximum $\langle Q_{sk} \rangle$ values of -45, -90, and -130 for the 60, 70, and 80 nm films, respectively. On sweeping into negative fields, the labyrinth domains fragment into skyrmions of opposite polarity, i.e., $\langle Q_{sk} \rangle$ = +1 (Figure 6(i,o,u)), with reduced total skyrmion numbers of +36, +65, and +90 for the respective thicknesses. Bloch-type helicity is stabilized because we have set lower DMI values. The complete field-dependent evolution of skyrmion from positive to negative saturation is provided for the 60 nm, 70 nm, and 80 nm GdFe films in Supplementary Movies 4–6.

Notably, the simulated $\langle Q_{sk} \rangle$ mirrors the experimental $\Delta \rho_{xy}$ trends (Figures 2–4), consistent with the proportionality $\Delta \rho_{xy} \propto \langle Q_{sk} \rangle$. As we have discussed, the characteristic double-peak feature arises from the sequential stabilization of two distinct skyrmion phases during the field sweep, as the core polarity reverses. Representative skyrmion textures at +0.25 T and -0.25 T are shown in the inset of Figure 6(a), where the skyrmion cores point outward and inward, corresponding to positive and negative $\langle Q_{sk} \rangle$, respectively. Both experimental data and simulations further show that $\Delta \rho_{xy}$ and $\langle Q_{sk} \rangle$ values increase with film thickness, reflecting the thickness-dependent enhancement of skyrmion density. Importantly, an asymmetry in $\langle Q_{sk} \rangle$ values emerges between positive and negative fields. This asymmetry increases with thickness and directly accounts for the experimentally observed differences in the magnitudes of the $\Delta \rho_{xy}$ peaks. 

\subsection{Thickness-Driven Control of Ferrimagnetic Skyrmions}
To establish a direct comparison between experiment and simulation, we analyzed the evolution of the skyrmion diameter ($d_{sk}$) and density ($n_{sk}$) as a function of film thickness. Figures 7a–c display the experimentally observed skyrmions in GdFe films of thickness 60, 70, and 80 nm, while Figures 7d–f present the corresponding simulated skyrmion configurations. All images are shown with a common scale bar of 0.5 $\mu m$, enabling a direct comparison of skyrmion density across thicknesses. Insets provide magnified views (0.05 $\mu m$ scale) of individual skyrmions, highlighting the variation in size and chirality with increasing thickness. 
\begin{figure*}[ht]
\centering
\includegraphics[width = 5.5in ,height= 3.1in]{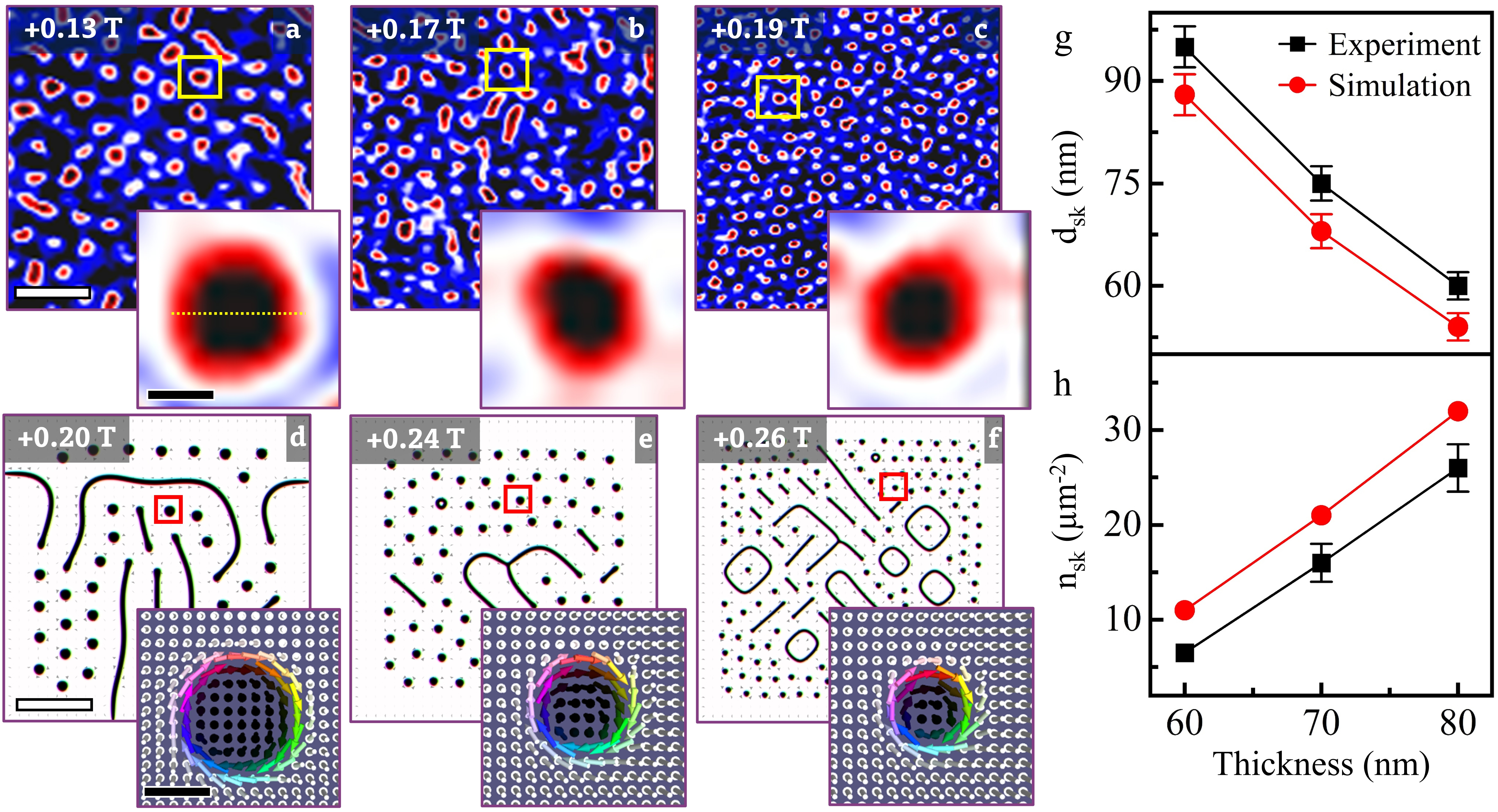}
\caption{Comparison of experimental and simulated skyrmion characteristics in single-layer GdFe films. MFM images of skyrmions in GdFe films with thicknesses of (a) 60 nm, (b) 70 nm, and (c) 80 nm, compared with corresponding micromagnetic simulations for (d) 60 nm, (e) 70 nm, and (f) 80 nm. Insets highlight individual skyrmions for enhanced visibility. All main panels share a scale bar of 0.5 $\mu m$ (shown in a and d), while inset images share a scale bar of 0.05 $\mu m$. (g) Extracted skyrmion diameter $d_{sk}$, and (h) Skyrmion density $n_{sk}$, as a function of film thickness.}
\end{figure*}
From the simulations, the skyrmion chirality can be unambiguously identified as Bloch-type. The spin configuration exhibits a characteristic vortex-like rotation, with spins oriented tangentially around the skyrmion's core. Specifically, the core spins align upward, while those at the periphery tilt downward, forming the hallmark Bloch rotation profile. A clear trend emerges in which the size and density of skyrmions evolve systematically with film thickness. The thickness dependence of the averaged skyrmion diameter is summarized in Figure 7g for both the experimental and simulated results. In both cases, the skyrmion diameter decreases monotonically with increasing thickness, reaching values of $\sim 60$ nm in the experiment and $\sim 55$ nm in the simulations. The skyrmion diameter in both the MFM and simulated images was quantified by applying a Gaussian fit to the line profiles (illustrated by the yellow dotted line in the inset of Fig. 7a), with the full width at half maximum (FWHM) taken as the characteristic size (See Figure S3 and S4). Figure 7h presents the quantitative comparison of skyrmion density extracted from experiment and simulation. Skyrmion density was determined by counting the number of skyrmions within 2 × 2 $\mu m^2$ regions of the MFM images using strict morphological criteria. Skyrmions were defined as closed, near-circular magnetic objects exhibiting a well-defined central contrast surrounded by an oppositely contrasted ring (as highlighted in Figure 7a-c), indicative of a full $2\pi$ spin rotation. Elongated stripe domains and magnetic bubbles lacking complete circular closure or showing direct connectivity to stripe domains were excluded from the analysis. Apart from these, density quantification was restricted to the high-field ranges, where isolated skyrmions were clearly stabilized and well separated, thereby avoiding strongly mixed-phase regimes. Furthermore, for each field value, skyrmions were counted within multiple, randomly selected 2 × 2 $\mu m^2$ regions across the image to ensure statistical reliability, and the reported densities represent averaged values. At 60 nm thickness, only a few skyrmions are observed, while at 80 nm, the density increases markedly, reaching $\sim$ 26 $\mu m^{-2}$ experimentally and $\sim$ 32 $\mu m^{-2}$ in the simulations. 

This close agreement between experiment and theory underscores that the size and density of skyrmions in GdFe single-layer films can be systematically tuned by altering the film thickness. This tunability arises from the interplay of key material parameters, such as $M_s$, $K_u$, A, and DMI strength. Experimentally, both $M_s$ and $K_u$ are found to increase with thickness. At the same time, simulations reveal that an enhanced bulk DMI with thickness is necessary to reproduce the observed skyrmion characteristics. Taken together, these effects jointly govern the evolution of skyrmions' size and density, which is generally described by a critical parameter $\kappa = \pi D/4\sqrt{AK_u}$ \cite{xu2023systematic,yu2016room}. Apart from these, in relatively thick single-layer films, the competition between DMI and the magnetic dipolar interactions further promotes the formation of dense skyrmion lattices \cite{erickson2024room,sivakumar2020topological}. This highlights the robustness of thickness-driven control over skyrmion properties, as demonstrated in our study, even though the composition gradient is reduced at higher film thicknesses.

We note that this study provides the first microscopic and electrical evidence of RT skyrmions in a FiM single-layer GdFe film, underscoring the versatility of MFM imaging and Hall transport as complementary probes for skyrmion detection in RT device-relevant conditions. The stabilization of RT skyrmions in our GdFe films can be attributed to three key factors. First, all films exhibit structural inversion asymmetry originating from a compositional gradient along the growth direction, which gives rise to a finite bulk DMI. Although the gradient decreases with increasing thickness, the DMI strength is not solely determined by asymmetry. Instead, it emerges from the delicate balance between anisotropy, exchange, and domain-wall energies, consistent with the relation $\sigma_{DW} = 4\sqrt{AK_u} - \pi |D|$ \cite{heide2008dzyaloshinskii,thiaville2012dynamics}. Since the experimentally extracted $K_u$ increases with film thickness, an increase in the effective DMI is physically reasonable and required to stabilize the observed magnetic skyrmions. Second, the intrinsic properties of the GdFe ferrimagnet play a pivotal role, such as the anisotropic antiferromagnetic coupling between the non-collinear Fe and Gd sublattices combined with the low net magnetization and weak perpendicular anisotropy arising from the vanishing orbital moment of Gd, naturally favor the stabilization of skyrmionic phases in weak-PMA regimes \cite{kim2017fast}. Third, the competition among PMA, enhanced dipolar interactions in thicker films, and bulk DMI promotes the stabilization of skyrmions with nonzero topological charge, where dipolar interactions support domain modulation while a finite DMI remains essential to select the chiral, topologically nontrivial skyrmion state \cite{birch2022history}. Compared with recent reports of gradient-engineered CoPt single layers \cite{erickson2025effect}, our GdFe FiM films offer distinct advantages. Specifically: (i) RT skyrmion stabilization in a FiM host, which is intrinsically advantageous over ferromagnets for low-power spintronic applications due to reduced net magnetization, suppressed skyrmion Hall effect, and faster spin dynamics \cite{kim2022ferrimagnetic}; (ii) a significantly larger topological Hall response ( $\sim$ 1.18 $\mu\Omega.cm$ compared to $\sim$ 0.25 $\mu\Omega.cm$) indicating a stronger emergent electromagnetic field associated with the skyrmion texture, confirming the stability and improved electrical detectability for devices; and (iii) a higher skyrmion density with smaller size. Together, these results broaden the material landscape for skyrmions and establish FiM single layers as a promising platform for future high-density, energy-efficient spintronic devices.

\section{Conclusion}
In this study, we have demonstrated thickness-driven control of ferrimagnetic skyrmions at room temperature in GdFe films with thicknesses of 60, 70, and 80 nm. By systematically tuning the film thickness, we achieved control over skyrmion stability, including a clear evolution of skyrmion properties: the skyrmion size decreases to $\sim 60$ nm. At the same time, the density increases up to $\sim$ 26 $\mu m^{-2}$. Electrical transport and MFM measurements reveal a strong correlation between the topological Hall resistivity and skyrmion density, which is further supported by micromagnetic simulations. A compositional gradient across the film thickness was observed via scanning transmission electron microscopy, indicating an asymmetric elemental distribution and potential inversion-symmetry breaking, which contribute to the emergence of bulk DMI. Interestingly, our results show that skyrmions can remain stable in thicker GdFe films where dipolar interactions become increasingly important, while a finite bulk DMI is still required to impart chirality and topological stability. This highlights the cooperative role of dipolar interactions, magnetic anisotropy, and bulk DMI in stabilizing skyrmions under appropriate magnetic conditions. We believe that our findings establish a viable route to engineer ferrimagnetic skyrmions with tunable size and density, which offers a promising opportunity for energy-efficient skyrmion-based spintronic devices. Looking ahead with the present work, a futuristic direction is to explore the dynamic control of these skyrmions, aiming to achieve high mobility and a reduced skyrmion Hall effect.

\section{Experimental Section}
\textbf{Sample Preparation.} Amorphous GdFe thin films with 60 nm, 70 nm, and 80 nm thicknesses were deposited on Si substrates using electron-beam evaporation under ultra-high vacuum conditions (base pressure $ < 2 \times 10^{-8}$ mbar). An equiatomic GdFe alloy target was employed, and the deposition was carried out at a controlled rate of 0.4 \AA/s. A 3 nm thin chromium capping layer was deposited in situ immediately after GdFe film deposition, without breaking the vacuum, to prevent oxidation. The substrates were continuously rotated at 20 rpm during deposition to ensure uniform film thickness.

\textbf{TEM observation.} 
Cross-sectional transmission electron microscopy observations were performed using a Titan G2 80-200 microscope equipped with a probe aberration corrector to investigate the depth-resolved microstructure of the GdFe films. Electron-transparent lamellae were prepared via a focused ion beam (FIB) lift-out technique (FEI Helios Nanolab 650). A protective nickel capping layer was deposited on the film surface prior to thinning to mitigate ion-beam-induced damage during FIB milling. Elemental analysis was conducted using energy-dispersive X-ray spectroscopy (EDS) with an FEI Super-X detector, and the resulting data were processed using Bruker ESPRIT software (version 1.9).

\textbf{Magnetic and Transport Measurement.} 
Magnetization reversal loops of the GdFe single-layer films were measured using a superconducting quantum interference device (SQUID) magnetometer (Quantum Design MPMS3) in both out-of-plane and in-plane sample orientations. Magnetotransport properties were characterized using a custom-built variable temperature insert (VTI) integrated into a high-field magnet system, complemented by a Quantum Design Physical Property Measurement System (PPMS). Hall measurements were performed using the van der Pauw geometry (See Figure S5) \cite{van1958method}, with the voltage probes positioned orthogonally to the current probes to ensure accurate detection of transverse voltage. The transverse Hall resistance $\rho_{xy}$ was recorded as a function of an OOP field. Hall coefficients were determined using a lock-in amplification technique (excitation frequency: 0–300 Hz), enabling sub-nanovolt resolution. All Hall measurements were performed over a complete hysteresis loop with a field step size of 10 mT, using a low dc excitation current of 200 $\mu A$  to minimize perturbation to the spin textures. All magnetic and transport data were acquired at 300 K, with identical field step increments to ensure precise extraction of the Hall signal. Notably, careful alignment and calibration procedures were employed to correct for magnetic field offsets between magnetization and transport measurements, providing accurate and reliable correlation between the two datasets.

\textbf{Magnetic Force Microscopy.}
Magnetic force microscopy (MFM) measurements were conducted using an NX-10 scanning probe microscope (Park Systems) equipped with a high-performance magnetic field generator (MFG, CAYLAR). The MFG allows precise application of both in-plane ($\pm$ 0.6 T) and out-of-plane (OOP) ($\pm$ 0.2 T) magnetic fields and enables fully automated sequential imaging under variable fields without affecting topography performance. The system maintains a low noise level ($<$50 pm) and exhibits no scan distortion or drift during field modulation. All measurements were performed under ambient conditions on a vibration-isolated platform. To minimize probe–sample interactions and suppress potential tip-induced perturbations, MFM measurements were performed with a low-moment magnetic probe (Bruker MESP-LM-V2) and a 40 nm lift height maintained during magnetic imaging. The estimated stray magnetic field of the MFM tip ($\sim$0.057 T) \cite{zhang2018direct} is substantially smaller than the magnetic field required to fully saturate the GdFe films ($\sim$0.3 T), thereby ruling out tip-induced modification of the magnetic textures. All MFM measurements were carried out in standard two-pass tapping mode, where the first pass records the surface topography and the second pass acquires the magnetic phase signal at a fixed lift height of 40 nm. Lift-height-dependent measurements (Figure S6) and repeated scans performed at a fixed lift height of 40 nm with different scan directions (Figure S7) revealed no discernible changes in domain size, shape, or spatial distribution. These observations confirm that the imaged domain and skyrmion morphologies are intrinsic to the GdFe films and are not influenced by MFM tip effects.

\textbf{Micromagnetic Simulations.}
Micromagnetic simulations were conducted to gain a deeper understanding of skyrmion dynamics using the $Mumax^3$ software package \cite{vansteenkiste2014design}. This software numerically solves the Landau-Lifshitz-Gilbert equation using a finite-difference approach \cite{abert2019micromagnetics} and incorporates bulk DMI interaction. The simulated geometry consisted of a $2 \times 2$ $\mu m^{2}$ film discretized into cells of $4 \times 4 \times t_{FiM}$ $nm^{3}$. The initial magnetization configuration was randomly oriented along the z direction perpendicular to the film surface, and allowed to relax to its ground state to minimize the system’s total magnetic energy. All simulations were conducted at 300 K. The input magnetic parameters used in the simulations were obtained from SQUID magnetometry measurements. The value of $M_s$ was set to 240, 310, and 380 kA/m for 60, 70, and 80 nm films, respectively, while the $K_u$ were taken as 50, 90, and 150 $kJ/m^3$ for the same thicknesses. The exchange stiffness constant A = 2.6 pJ/m is adopted from the literature \cite{miguel2006x}. We systematically varied the bulk DMI values (0.4, 0.5, and 0.6 $mJ/m^2$) for film thicknesses of 60, 70, and 80 nm to compare the field-dependent evolution of skyrmion size and density with experimental results.

\section*{Supporting Information}
Details of the Micromagnetic simulation, skyrmion line-scan profiles obtained from the MFM and simulation, a schematic configuration of the Hall resistivity measurements, and lift-height-dependent measurements were presented. Field-dependent evolution of skyrmions obtained from MFM measurements for 60 nm (Movie 1), 70 nm (Movie 2), and 80 nm (Movie 3) GdFe films, and from micromagnetic simulations for 60 nm (Movie 4), 70 nm (Movie 5), and 80 nm (Movie 6).

\section*{Acknowledgements}
S.K.M. acknowledges IIT Hyderabad and NIMS, Japan, for providing research facilities, as well as the University Grants Commission (UGC, India) for providing financial support. J.R.M. acknowledges funding support from DST Nanomission (project no.: DST/NM/TUE/QM-4/2019-IIT-H). This work was also supported by intramural funding at TIFR-Hyderabad from the Department of Atomic Energy, Government of India, under Project Identification No. RTI 4007, and external funding from ONRG Grant No. N62909-23-1-2049. The authors thank Ms. Yukie Mori (National Institute for Materials Science) for FIB sample preparation.

\section*{Conflict of interest}
The authors declare no conﬂict of interest.

\section*{Data Availability Statement}
The data supporting the findings of this study are available from the corresponding author upon reasonable request.

\printendnotes

% Submissions are not required to reflect the precise reference formatting of the journal (use of italics, bold etc.), however it is important that all key elements of each reference are included.
%\bibliography{sample}

\begin{thebibliography}{}
\bibitem{fert2013skyrmions}
Fert, A., Cros, V. and Sampaio, J., 2013. Skyrmions on the track. Nat. Nanotechnol., 8(3), pp.152-156.
\bibitem{everschor2018perspective}
Everschor-Sitte, K., Masell, J., Reeve, R.M. and Kläui, M., 2018. Perspective: Magnetic skyrmions—Overview of recent progress in an active research field. J. Appl. Phys., 124(24).
\bibitem{nagaosa2013topological}
Nagaosa, N. and Tokura, Y., 2013. Topological properties and dynamics of magnetic skyrmions. Nat. Nanotechnol., 8(12), pp.899-911.
\bibitem{fert2017magnetic}
Fert, A., Reyren, N. and Cros, V., 2017. Magnetic skyrmions: advances in physics and potential applications. Nat. Rev. Mater., 2(7), pp.1-15.
\bibitem{hagemeister2015stability}
Hagemeister, J., Romming, N., Von Bergmann, K., Vedmedenko, E.Y. and Wiesendanger, R., 2015. Stability of single skyrmionic bits. Nat. Commun., 6(1), p.8455.
\bibitem{parkin2008magnetic}
Parkin, S.S., Hayashi, M. and Thomas, L., 2008. Magnetic domain-wall racetrack memory. Sci., 320(5873), pp.190-194.
\bibitem{zhang2015magnetic}
Zhang, X., Ezawa, M. and Zhou, Y., 2015. Magnetic skyrmion logic gates: conversion, duplication and merging of skyrmions. Sci. Rep., 5(1), pp.1-8.
\bibitem{zhang2015magnetics}
Zhang, X., Zhou, Y., Ezawa, M., Zhao, G.P. and Zhao, W., 2015. Magnetic skyrmion transistor: skyrmion motion in a voltage-gated nanotrack. Sci. Rep., 5(1), p.11369.
\bibitem{song2020skyrmion}
Song, K.M., Jeong, J.S., Pan, B., Zhang, X., Xia, J., Cha, S., Park, T.E., Kim, K., Finizio, S., Raabe, J. and Chang, J., 2020. Skyrmion-based artificial synapses for neuromorphic computing. Nat. Electron., 3(3), pp.148-155.
\bibitem{finocchio2016magnetic}
Finocchio, G., Büttner, F., Tomasello, R., Carpentieri, M. and Kläui, M., 2016. Magnetic skyrmions: from fundamental to applications. J. Phys. D: Appl. Phys., 49(42), p.423001.
\bibitem{bode2007chiral}
Bode, M., Heide, M., Von Bergmann, K., Ferriani, P., Heinze, S., Bihlmayer, G., Kubetzka, A., Pietzsch, O., Blügel, S. and Wiesendanger, R., 2007. Chiral magnetic order at surfaces driven by inversion asymmetry. Nat., 447(7141), pp.190-193.
\bibitem{heide2008dzyaloshinskii}
Heide, M., Bihlmayer, G. and Blügel, S., 2008. Dzyaloshinskii-Moriya interaction accounting for the orientation of magnetic domains in ultrathin films: Fe/W (110). Phys. Rev. B - Condens. Matter Mater. Phys., 78(14), p.140403.
\bibitem{dzyaloshinsky1958thermodynamic}
Dzyaloshinsky, I., 1958. A thermodynamic theory of “weak” ferromagnetism of antiferromagnetics. J. Phys. Chem. Solids, 4(4), pp.241-255.
\bibitem{moriya1960anisotropic}
Moriya, T., 1960. Anisotropic superexchange interaction and weak ferromagnetism. Phys. Rev., 120(1), p.91.
\bibitem{lee2009unusual}
Lee, M., Kang, W., Onose, Y., Tokura, Y. and Ong, N.P., 2009. Unusual Hall effect anomaly in MnSi under pressure. Phys. Rev. Lett., 102(18), p.186601.
\bibitem{kanazawa2011large}
Kanazawa, N., Onose, Y., Arima, T., Okuyama, D., Ohoyama, K., Wakimoto, S., Kakurai, K., Ishiwata, S. and Tokura, Y., 2011. Large topological Hall effect in a short-period helimagnet MnGe. Phys. Rev. Lett., 106(15), p.156603.
\bibitem{muhlbauer2009skyrmion}
Muhlbauer, S., Binz, B., Jonietz, F., Pfleiderer, C., Rosch, A., Neubauer, A., Georgii, R. and Boni, P., 2009. Skyrmion lattice in a chiral magnet. Sci., 323(5916), pp.915-919.
\bibitem{soumyanarayanan2017tunable}
Soumyanarayanan, A., Raju, M., Gonzalez Oyarce, A.L., Tan, A.K., Im, M.Y., Petrović, A.P., Ho, P., Khoo, K.H., Tran, M., Gan, C.K., and Ernult, F., 2017. Tunable room-temperature magnetic skyrmions in Ir/Fe/Co/Pt multilayers. Nat. Mater., 16(9), pp.898-904.
\bibitem{woo2016observation}
Woo, S., Litzius, K., Krüger, B., Im, M.Y., Caretta, L., Richter, K., Mann, M., Krone, A., Reeve, R.M., Weigand, M. and Agrawal, P., 2016. Observation of room-temperature magnetic skyrmions and their current-driven dynamics in ultrathin metallic ferromagnets. Nat. Mater., 15(5), pp.501-506.
\bibitem{pollard2017observation}
Pollard, S.D., Garlow, J.A., Yu, J., Wang, Z., Zhu, Y., and Yang, H., 2017. Observation of stable Néel skyrmions in cobalt/palladium multilayers with Lorentz transmission electron microscopy. Nat. Commun., 8(1), p.14761.
\bibitem{erickson2024room}
Erickson, A., Zhang, Q., Vakili, H., Li, C., Sarin, S., Lamichhane, S., Jia, L., Fescenko, I., Schwartz, E., Liou, S.H., and Shield, J.E., 2024. Room Temperature Magnetic Skyrmions in Gradient-Composition Engineered CoPt Single Layers. ACS nano, 18(45), pp.31261-31273.
\bibitem{litzius2017skyrmion}
Litzius, K., Lemesh, I., Krüger, B., Bassirian, P., Caretta, L., Richter, K., Büttner, F., Sato, K., Tretiakov, O.A., Förster, J. and Reeve, R.M., 2017. Skyrmion Hall effect revealed by direct time-resolved X-ray microscopy. Nat. Phys., 13(2), pp.170-175.
\bibitem{woo2018current}
Woo, S., Song, K.M., Zhang, X., Zhou, Y., Ezawa, M., Liu, X., Finizio, S., Raabe, J., Lee, N.J., Kim, S.I. and Park, S.Y., 2018. Current-driven dynamics and inhibition of the skyrmion Hall effect of ferrimagnetic skyrmions in GdFeCo films. Nat. Commun., 9(1), p.959.
\bibitem{caretta2018fast}
Caretta, L., Mann, M., Büttner, F., Ueda, K., Pfau, B., Günther, C.M., Hessing, P., Churikova, A., Klose, C., Schneider, M. and Engel, D., 2018. Fast current-driven domain walls and small skyrmions in a compensated ferrimagnet. Nat. Nanotechnol., 13(12), pp.1154-1160.
\bibitem{xu2023systematic}
Xu, T., Zhang, Y., Wang, Z., Bai, H., Song, C., Liu, J., Zhou, Y., Je, S.G., N’Diaye, A.T., Im, M.Y. and Yu, R., 2023. Systematic Control of Ferrimagnetic Skyrmions via Composition Modulation in Pt/Fe1–x Tb x/Ta Multilayers. ACS nano, 17(8), pp.7920-7928.
\bibitem{yang2015anatomy}
Yang, H., Thiaville, A., Rohart, S., Fert, A. and Chshiev, M., 2015. Anatomy of Dzyaloshinskii-Moriya interaction at Co/Pt interfaces. Phys. Rev. Lett., 115(26), p.267210.
\bibitem{belmeguenai2018thickness}
Belmeguenai, M., Roussigné, Y., Bouloussa, H., Chérif, S.M., Stashkevich, A., Nasui, M., Gabor, M.S., Mora-Hernández, A., Nicholson, B., Inyang, O.O. and Hindmarch, A.T., 2018. Thickness dependence of the Dzyaloshinskii-Moriya interaction in Co 2 FeAl Ultrathin films: Effects of annealing temperature and heavy-metal material. Phys. Rev. Appl., 9(4), p.044044.
\bibitem{buttner2018theory}
Buttner, F., Lemesh, I. and Beach, G.S., 2018. Theory of isolated magnetic skyrmions: From fundamentals to room temperature applications. Sci. Rep., 8(1), p.4464.
\bibitem{siemens2016minimal}
Siemens, A., Zhang, Y., Hagemeister, J., Vedmedenko, E.Y. and Wiesendanger, R., 2016. Minimal radius of magnetic skyrmions: statics and dynamics. New J. Phys., 18(4), p.045021.
\bibitem{meersschaut1995spin}
Meersschaut, J., Dekoster, J., Schad, R., Beliën, P. and Rots, M., 1995. Spin density wave instability for chromium in Fe/Cr (100) multilayers. Phys. Rev. Lett., 75(8), p.1638.
\bibitem{zabel1999magnetism}
Zabel, H., 1999. Magnetism of chromium at surfaces, at interfaces and in thin films. J. Phys. Condens. Matter, 11(48), p.9303.
\bibitem{kim2019bulk}
Kim, D.H., Haruta, M., Ko, H.W., Go, G., Park, H.J., Nishimura, T., Kim, D.Y., Okuno, T., Hirata, Y., Futakawa, Y. and Yoshikawa, H., 2019. Bulk Dzyaloshinskii–Moriya interaction in amorphous ferrimagnetic alloys. Nat. Mater., 18(7), pp.685-690.
\bibitem{krishnia2021spin}
Krishnia, S., Haltz, E., Berges, L., Aballe, L., Foerster, M., Bocher, L., Weil, R., Thiaville, A., Sampaio, J. and Mougin, A., 2021. Spin-orbit coupling in single-layer ferrimagnets: direct observation of spin-orbit torques and chiral spin textures. Phys. Rev. Appl., 16(2), p.024040.
\bibitem{kumar2024engineering}
Kumar Mishra, S., Prasanth Perumal, H. and Mohanty, J., 2024. Engineering perpendicular magnetic anisotropy and Dzyaloshinskii–Moriya interaction in Gd–Fe thin films for spintronics applications. J. Appl. Phys., 136(24).
\bibitem{mallick2024driving}
Mallick, S., Sassi, Y., Prestes, N.F., Krishnia, S., Gallego, F., M. Vicente Arche, L., Denneulin, T., Collin, S., Bouzehouane, K., Thiaville, A. and Dunin-Borkowski, R.E., 2024. Driving skyrmions in flow regime in synthetic ferrimagnets. Nat. Commun., 15(1), p.8472.
\bibitem{harris1992structural}
Harris, V.G., Aylesworth, K.D., Das, B.N., Elam, W.T. and Koon, N.C., 1992. Structural origins of magnetic anisotropy in sputtered amorphous Tb-Fe films. Phys. Rev. Lett., 69(13), p.1939.
\bibitem{hansen1989magnetic}
Hansen, P., Clausen, C., Much, G., Rosenkranz, M. and Witter, K., 1989. Magnetic and magneto‐optical properties of rare‐earth transition‐metal alloys containing Gd, Tb, Fe, Co. J. Appl. Phys., 66(2), pp.756-767.
\bibitem{heigl2021dipolar}
Heigl, M., Koraltan, S., Vaňatka, M., Kraft, R., Abert, C., Vogler, C., Semisalova, A., Che, P., Ullrich, A., Schmidt, T. and Hintermayr, J., 2021. Dipolar-stabilized first and second-order antiskyrmions in ferrimagnetic multilayers. Nat. Commun., 12(1), p.2611.
\bibitem{legrand2018hybrid}
Legrand, W., Chauleau, J.Y., Maccariello, D., Reyren, N., Collin, S., Bouzehouane, K., Jaouen, N., Cros, V. and Fert, A., 2018. Hybrid chiral domain walls and skyrmions in magnetic multilayers. Sci. Adv., 4(7), p.eaat0415.
\bibitem{kim2017fast}
Kim, K.J., Kim, S.K., Hirata, Y., Oh, S.H., Tono, T., Kim, D.H., Okuno, T., Ham, W.S., Kim, S., Go, G. and Tserkovnyak, Y., 2017. Fast domain wall motion in the vicinity of the angular momentum compensation temperature of ferrimagnets. Nat. Mater., 16(12), pp.1187-1192.
\bibitem{kim2022ferrimagnetic}
Kim, S.K., Beach, G.S., Lee, K.J., Ono, T., Rasing, T. and Yang, H., 2022. Ferrimagnetic spintronics. Nat. Mater., 21(1), pp.24-34.
\bibitem{zhang2018determination}
Zhang, S., Zhang, J., Wen, Y., Chudnovsky, E.M. and Zhang, X., 2018. Determination of chirality and density control of Néel-type skyrmions with in-plane magnetic field. Commun. Phys., 1(1), p.36.
\bibitem{bhatti2023enhancement}
Bhatti, S., Tan, H.K., Sim, M.I., Zhang, V.L., Sall, M., Xing, Z.X., Juge, R., Mahendiran, R., Soumyanarayanan, A., Lim, S.T. and Ravelosona, D., 2023. Enhancement of skyrmion density via interface engineering. APL Mater., 11(1).
\bibitem{malozemoff1979magnetic}
Malozemoff, A.P. and Slonczewski, J.C., 1979. Magnetic domain walls in bubble materials, Appl. Solid State Sci.(Academic Press, New York, NY, 1979).
\bibitem{yu2016room}
Yu, G., Upadhyaya, P., Li, X., Li, W., Kim, S.K., Fan, Y., Wong, K.L., Tserkovnyak, Y., Amiri, P.K. and Wang, K.L., 2016. Room-temperature creation and spin–orbit torque manipulation of skyrmions in thin films with engineered asymmetry. Nano Lett., 16(3), pp.1981-1988.
\bibitem{zhang2018creation}
Zhang, S., Zhang, J., Wen, Y., Chudnovsky, E.M. and Zhang, X., 2018. Creation of a thermally assisted skyrmion lattice in Pt/Co/Ta multilayer films. Appl. Phys. Lett., 113(19).
\bibitem{mishra2017anomalous}
Mishra, R., Yu, J., Qiu, X., Motapothula, M., Venkatesan, T. and Yang, H., 2017. Anomalous current-induced spin torques in ferrimagnets near compensation. Phys. Rev. Lett., 118(16), p.167201.
\bibitem{talapatra2019observation}
Talapatra, A., Chelvane, J.A. and Mohanty, J., 2019. Observation of magnetic domains in Gd-Fe thin films with complementary microscopy techniques. J. Magn. Magn. Mater., 489, p.165469.
\bibitem{mishra2025exploring}
Mishra, S.K., Poddar, N.P., Chelvane, J.A. and Mohanty, J., 2025. Exploring thickness effects on magnetization reversal mechanism and domain state configuration in ferrimagnetic Gd–Fe thin film. J. Mater. Sci.: Mater. Electron., 36(7), p.413.
\bibitem{raju2021colossal}
Raju, M., Petrović, A.P., Yagil, A., Denisov, K.S., Duong, N.K., Göbel, B., Şaşıoğlu, E., Auslaender, O.M., Mertig, I., Rozhansky, I.V. and Panagopoulos, C., 2021. Colossal topological Hall effect at the transition between isolated and lattice-phase interfacial skyrmions. Nat. Commun., 12(1), p.2758.
\bibitem{neubauer2009topological}
Neubauer, A., Pfleiderer, C., Binz, B., Rosch, A., Ritz, R., Niklowitz, P.G. and Böni, P., 2009. Topological Hall effect in the A phase of MnSi. Phys. Rev. Lett., 102(18), p.186602.
\bibitem{sivakumar2020topological}
Sivakumar, P.K., Gobel, B., Lesne, E., Markou, A., Gidugu, J., Taylor, J.M., Deniz, H., Jena, J., Felser, C., Mertig, I. and Parkin, S.S., 2020. Topological Hall signatures of two chiral spin textures hosted in a single tetragonal inverse Heusler thin film. ACS nano, 14(10), pp.13463-13469.
\bibitem{jefremovas2025role}
Jefremovas, E.M., Leutner, K., Fischer, M.G., Marqués-Marchán, J., Winkler, T.B., Asenjo, A., Sinova, J., Frömter, R. and Kläui, M., 2025. The role of magnetic dipolar interactions in skyrmion lattices. Newton, 1(2).
\bibitem{ojha2025tailoring}
Ojha, B., Mohanty, S., Sharma, B. and Bedanta, S., 2025. Tailoring the topological Hall effect in Pt/Co/X (X= Ta, Re) thin films. Phys. Rev. B., 112(6), p.064419.
\bibitem{tai2022distinguishing}
Tai, L., Dai, B., Li, J., Huang, H., Chong, S.K., Wong, K.L., Zhang, H., Zhang, P., Deng, P., Eckberg, C. and Qiu, G., 2022. Distinguishing the two-component anomalous Hall effect from the topological Hall effect. ACS nano, 16(10), pp.17336-17346.
\bibitem{denisov2020theory}
Denisov, K.S., 2020. Theory of an electron asymmetric scattering on skyrmion textures in two-dimensional systems. J. Phys. Condens. Matter., 32(41), p.415302.p.267201.
\bibitem{vansteenkiste2014design}
Vansteenkiste, A., Leliaert, J., Dvornik, M., Helsen, M., Garcia-Sanchez, F. and Van Waeyenberge, B., 2014. The design and verification of MuMax3. AIP Adv., 4(10).
\bibitem{thiaville2012dynamics}
Thiaville, A., Rohart, S., Jué, É., Cros, V. and Fert, A., 2012. Dynamics of Dzyaloshinskii domain walls in ultrathin magnetic films. Europhys. Lett., 100(5), p.57002.
\bibitem{birch2022history}
Birch, M.T., Powalla, L., Wintz, S., Hovorka, O., Litzius, K., Loudon, J.C., Turnbull, L.A., Nehruji, V., Son, K., Bubeck, C. and Rauch, T.G., 2022. History-dependent domain and skyrmion formation in 2D van der Waals magnet Fe3GeTe2. Nat. Commun., 13(1), p.3035.
\bibitem{erickson2025effect}
Erickson, A., Zhang, Q., Vakili, H., Schwartz, E., Lamichhane, S., Li, C., Li, B., Song, D., Chai, G., Liou, S.H. and Kovalev, A.A., 2025. Effect of Magnetic Anisotropy and Gradient‐Induced Dzyaloshinskii‐Moriya Interaction on the Formation of Magnetic Skyrmions. Small, p.e05204.
\bibitem{van1958method}
van der Pauw, L.J., 1958. A method of measuring the resistivity and Hall coefficient on lamellae of arbitrary shape. PHILIPS TECH. REV., 20, pp.220-224.
\bibitem{zhang2018direct}
Zhang, S., Zhang, J., Zhang, Q., Barton, C., Neu, V., Zhao, Y., Hou, Z., Wen, Y., Gong, C., Kazakova, O. and Wang, W., 2018. Direct writing of room temperature and zero field skyrmion lattices by a scanning local magnetic field. Appl. Phys. Lett., 112(13).
\bibitem{abert2019micromagnetics}
Abert, C., 2019. Micromagnetics and spintronics: models and numerical methods. Eur. Phys. J. B., 92(6), p.120.
\bibitem{miguel2006x}
Miguel, J., Peters, J.F., Toulemonde, O.M., Dhesi, S.S., Brookes, N.B. and Goedkoop, J.B., 2006. X-ray resonant magnetic scattering study of magnetic stripe domains in a-GD FE thin films. Phys. Rev. B - Condens. Matter Mater. Phys., 74(9), p.094437.
\end{thebibliography}

%\begin{biography}[example-image-1x1]{A.~One}

%\end{biography}

%\graphicalabstract{example-image-1x1}{Please check the journal's author guildines for whether a graphical abstract, key points, new findings, or other items are required for display in the Table of Contents.}

\end{document}